\documentclass[reprint,aps,prb,twocolumn,superscriptaddress,
floatfix,showpacs,amsmath,amssymb]{rjparticle}

\usepackage{verbatim}
\usepackage{graphicx,color,bbm}% Include figure files
\usepackage{epsfig}
\usepackage{enumerate}
\newcommand{\ketbra}[2]{\vert {#1} \rangle \langle{#2}\vert}
\newcommand{\grayeq}[1]{\textcolor[rgb]{0.7,0.7,0.7}{#1}}

\newcommand{\tcred}[1]{\textcolor[rgb]{0.7,0,0}{#1}}
\newcommand{\tcblue}[1]{\textcolor[rgb]{0,0,0.7}{#1}}

\newcommand{\tcgree}[1]{\textcolor[rgb]{0,0.6,0}{#1}}

\begin{document}
\title{Design of a Lambda configuration in artificial coherent 
nanostructures}

\author[1]{P.G. Di Stefano}
\author[1,2,3]{E. Paladino}
\author[1,2]{A. D'Arrigo}
\author[4,3]{B. Spagnolo}
\author[1,2,3]{G. Falci}
\affil[1]{Dipartimento di Fisica e Astronomia,
Universit\`a di Catania, Via Santa Sofia 64, 95123 Catania, Italy.}
\affil[2]{CNR-IMM  UOS Universit\`a (MATIS), 
Consiglio Nazionale delle Ricerche, Via Santa Sofia 64, 95123 Catania, Italy.}
\affil[3]{Istituto Nazionale di Fisica Nucleare, Via Santa Sofia 64, 95123 Catania, Italy.}
\affil[4]{Dipartimento di Fisica e Chimica, 
Universit\`{a} di Palermo, Group of Interdisciplinary Physics and CNISM,
Unit\`{a} di Palermo, Viale delle Scienze, Ed.18, I-90128 Palermo, Italy.}
\date{\today}
\pacs{03.67.Lx, 42.50.Gy, 03.65.Yz,  85.25.-j}
% 03.65.Yz Decoherence; open systems; quantum statistical methods
% 03.67.Lx Quantum computation architectures and implementations
% 03.67.Hk Quantum communication
% 89.70.+c Information theory and communication theory 
% 03.67.Pp Quantum error correction and other methods for protection against decoherence
% 42.50.Gy Effects of atomic coherence on propagation, absorption, and amplification of light; electromagnetically induced transparency and absorption
% 85.25.-j Superconducting devices
\begin{abstract}
The implementation of a three-level Lambda System 
in artificial atoms would allow to perform advanced 
control tasks typical of quantum optics in the 
solid state realm, with 
photons in the  $\mathrm{\mu m}$/mm range.
However hardware constraints put an obstacle since
protection from decoherence is often conflicting with 
efficient coupling to external fields.
We address the problem of performing conventional 
STImulated Raman Adiabatic 
Passage (STIRAP) in the presence of low-frequency noise. 
We propose two strategies to defeat decoherence, 
based on ``optimal symmetry breaking'' and dynamical 
decoupling. We suggest
how to apply to the different implementations of 
superconducting artificial atoms, stressing the key role of non-Markovianity.
%We also discuss two new unconventional schemes to bypass hardware constraints allowing to perform STIRAP at protected symmetry points and for always-on fields. 
\end{abstract}
\pacs{03.67.Lx,85.25.-j, 03.65.Yz}

\maketitle
\section{Introduction}
\label{sec.intro}
In recent years several experiments have demonstrated 
multilevel coherence in superconducting 
artificial atoms, as the observation 
of the Autler-Townes (AT)~\cite{ka:209-sillanpa-simmonds-prl-autlertownes,ka:212-li-hakonen-srep-qswitch} effect, 
 of electromagnetically induced transparency (EIT)~\cite{ka:210-abdumalikov-prl-eit}, besides evidences of three-state 
superpositions~\cite{ka:210-bianchetti-prl} and 
coherent population trapping 
(CPT)~\cite{ka:210-kellypappas-prl-cpt}. 
Further exploiting coherence 
in such systems would be important in principle and 
moreover allow important applications in solid-state 
quantum integrated coherent architectures.
So far all the experiments in these systems 
(except the one of Ref. ~\cite{ka:210-kellypappas-prl-cpt}) have been performed driving 
by ac-fields in {\em ladder} 
configuration (see Fig.~\ref{fig:stirap-ideal}a). 
In this work we address the 
design of a {\em lambda} configuration 
in three-level artificial atoms which would allow 
to implement tasks~\cite{ka:199-kuhn-APB-singlephoton,ka:213-muckerempe-pra-stirapsinglephgen,ka:209-siebrafalci-optcomm} 
where two-photon absorption {\em and} emission are invoved 
at once.
Despite of several theoretical proposal~\cite{ka:205-liunori-prl-adiabaticpassage,ka:208-weinori-prl-stirapqcomp,ka:212-falci-physscr,ka:213-falci-prb-stirapcpb,kr:211-younori-nature-multilevel}, 
this goal is still experimentally unsettled, mainly 
because protection from low-frequency noise requires to 
enforce exact or approximate symmetries of the Hamiltonian, which
on the other hand imply selection rules canceling the pump 
coupling~\cite{ka:205-liunori-prl-adiabaticpassage,ka:209-siebrafalci-optcomm,ka:213-falci-prb-stirapcpb}
(see Fig.~\ref{fig:stirap-ideal}a).  

Our second goal of is to elucidate the central role of 
non-Markovian noise in producing three-level decoherence 
for the class of phenomena based on CPT. We focus on 
a protocol called STIRAP~\cite{kr:198-bergmann-rmp-stirap,kr:201-vitanov-annurev}, 
described in Sec.\ref{sec:STIRAP}, 
which involves several basic coherent effects and 
allows striking applications in integrated atom-cavity systems.  
Therefore its demonstration would be a benchmark for 
multilevel advanced control in artificial atoms. 
In Sec.~\ref{sec:model-noise} we introduce an effective 
model for noise  and argue that dephasing in 
the ``trapped subspace'' $\mathrm{span}\{\ket{0},\ket{1}\}$
(see Fig.~\ref{fig:stirap-ideal}a) plays the major role. 
We show that implementation of STIRAP 
in Lambda configuration is possible within present technology. 
In Sec.~\ref{sec:figure-merit} we propose two strategies 
to defeat dephasing, namely the search for optimal 
symmetry breaking conditions, and selective dynamical 
decoupling of noise sources achieved by operating on 
a specific external control. 
Both strategies leverage on the fact that 
dephasing in the solid state is due to broad band colored 
noise (BBCN), which is inherently non-Markovian. 
As a consequence BBCN
impacts on dephasing in a way specific of 
correlations of the induced fluctuations of the device 
bandstructure.
\begin{comment}
In Sec.~\ref{sec:STIRAP21} we present two new schemes 
for the STIRAP control problem, coping with the constraints 
to control imposed by quantum hardware in artificial atoms 
and possibly allowing manipulation of photons in the 
in the  $\mathrm{\mu m}$/mm range. 
\end{comment}
Finally in 
Sec.\ref{sec:conclusions} we conclude and discuss some further 
perspective.

\section{Coherent population transfer in three-level atoms}
\label{sec:STIRAP}
STIRAP is an advanced control technique 
of $M>2$-level systems, 
allowing complete population transfer 
between two states $\ket{0}$ and $\ket{1}$, 
even in absence of a direct coupling, via 
one or more intermediate states which 
are {\em never} populated.  
\begin{figure}[t]
\centering
\includegraphics[width=0.27\columnwidth]{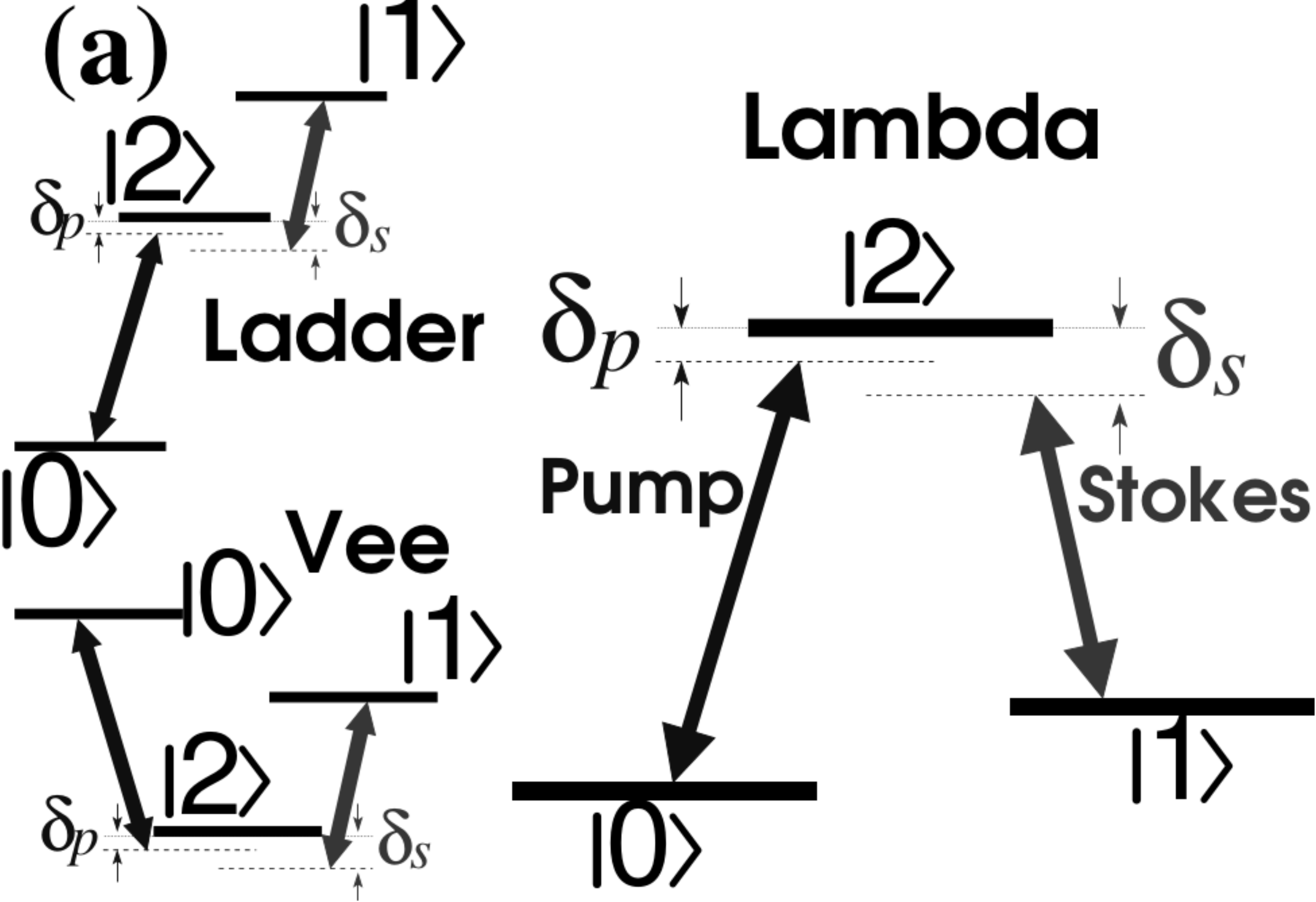}\quad
\includegraphics[width=0.27\columnwidth]{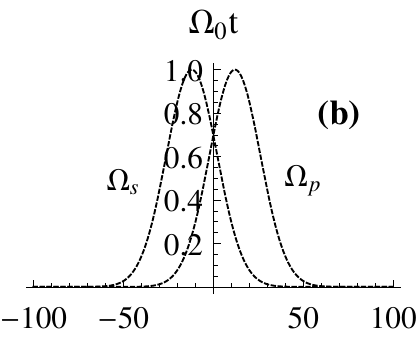}
\quad
\includegraphics[width=0.27\columnwidth]{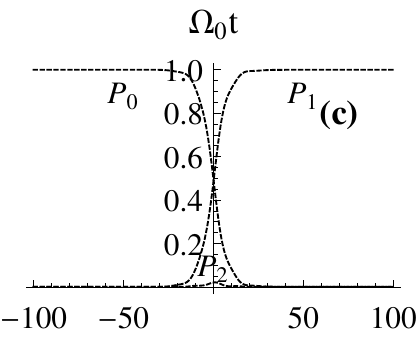}
\caption{(a) Three-level system driven with AC 
fields in $\Lambda$ configuration (in the insets the Ladder and 
the Vee configurations). 
(b) Gaussian pulses in the counterintuitive sequence (here $\Omega_0 T = 20, \tau= 0.6 \,T$).
(c) Population histories 
$P_{i}(t) = |\langle i |\psi(t) \rangle|^2$
for ideal STIRAP ($\delta =0$) and for 
$\delta_p= -0.2\,\Omega_0$,
$\kappa = 1$.
\label{fig:stirap-ideal}}
\end{figure}
In three-level systems the indirect linkage is provided by 
the typical configurations of two ac-fields 
shown in Fig.~\ref{fig:stirap-ideal}a. 
The pump field at $\omega_p \approx |E_2-E_0|$, 
triggers transitions $\ket{0} \leftrightarrow \ket{2}$ 
whereas the Stokes, $\omega_s \approx |E_2-E_1|$, triggers 
$\ket{1} \leftrightarrow \ket{2}$ ones.
The standard Hamiltonian in the rotating wave approximation
(RWA) in a rotating frame referred to 
the ``bare'' basis $\{\ket{0},\ket{1},\ket{2}\}$ is given  by the matrix
\begin{equation}
\label{eq:H}
H = \begin{bmatrix}
0 								& 	0 							& \frac{1}{2}\Omega^\ast_p(t)	\\
0 								& 	\delta (t)					& \frac{1}{2}\Omega^\ast_s(t) 	\\
\frac{1}{2}\Omega_p(t) 		& 	\frac{1}{2}\Omega_s(t) 	& \delta_p(t) 						\\
\end{bmatrix}
\end{equation}
where the Rabi frequencies $\Omega_k(t)$ for $k=p,s$ 
are related to the amplitudes of the pump and Stokes fields, 
$\delta_k$ are the single-photon detunings and
$\delta = \delta_p - \delta_s$ is the two-photon detuning.  
We will mostly refer to the Lambda configurations where 
$\delta_p(t):=E_2-E_0-\omega_p$ and  
$\delta_s(t):=E_2-E_1-\omega_s$.
At two-photon resonance, $\delta=0$, 
the Hamiltonian (\ref{eq:H}) has an instantaneous eigenvector 
with null eigenvalue, $\epsilon_0=0$, given by
$\ket{D}= (\Omega_s \ket{0} - \Omega_p \ket{1})
/\sqrt{\Omega_s^2+\Omega_p^2}$. 
%\begin{equation}
%\label{eq:dark.state}
%\ket{D}=\frac{\Omega_s \ket{0} - \Omega_p \ket{1}}{\sqrt{\Omega_s^2+\Omega_p^2}}$
%\end{equation}
It is called the ``dark state'' since state 
$\ket{2}$ is not populated, despite of the transitions triggered by the fields.  
%The other eigenstates $\ket{\pm}$ have eigenvalues $\epsilon_{\pm}= \frac{1}{2}\delta_p \pm \frac{1}{2} \Omega_{AT}$ where $\Omega_{AT} = \sqrt{\Omega_p^2 + \Omega_s^2 + \delta_p^2}$ is the AT splitting. 
In ideal STIRAP ($\delta=0$) 
adiabatic pulses $\Omega_k(t)$ are
shined in the 
\textit{counterintuitive} sequence, i.e. the Stokes preceding the pump. We will make use of Gaussian pulses
\begin{equation}
\Omega_p=\kappa_p \Omega_0 \,\mathrm{e}^{-[({t+\tau})/{T}]^2}\qquad \Omega_s=\kappa_s \Omega_0 \,\mathrm{e}^{-[({t-\tau}/{T}]^2}
\end{equation} 
with $\tau \sim T$. Here $\Omega_0$ is a frequency scale 
and $\kappa_k \sim 1$ are constants which will be taken equal 
to $1$ when not otherwise specified. 
In this way the dark state $\ket{D(t)}$
(Fig.~\ref{fig:stirap-ideal}b) performs the desired
$\ket{0} \to \ket{1}$ evolution, 
yielding complete population transfer, while $\ket{2}$
is never populated (Fig.~\ref{fig:stirap-ideal}d).

\begin{figure}[t]
\centering
\begin{minipage}[c]{0.25\columnwidth}
\includegraphics[width=0.8\columnwidth]{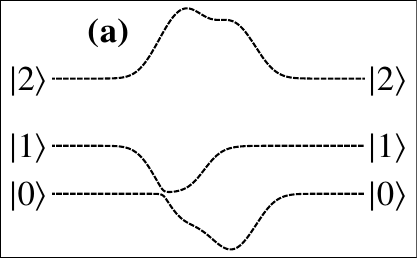}
\\
\includegraphics[width=0.8\columnwidth]{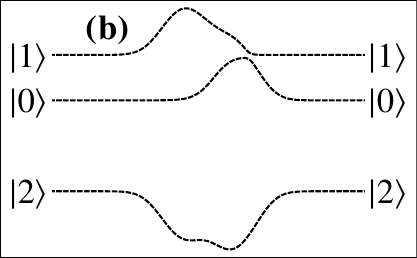}
\\
\includegraphics[width=0.8\columnwidth]{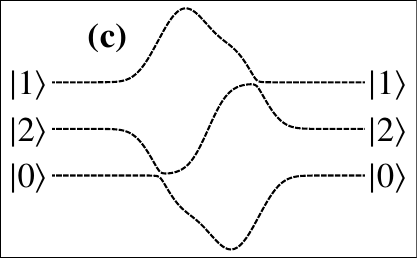}
\\
\end{minipage}
\begin{minipage}[c]{0.25\columnwidth}
%\begin{minipage}[t]{\columnwidth}
\includegraphics[width=\columnwidth]{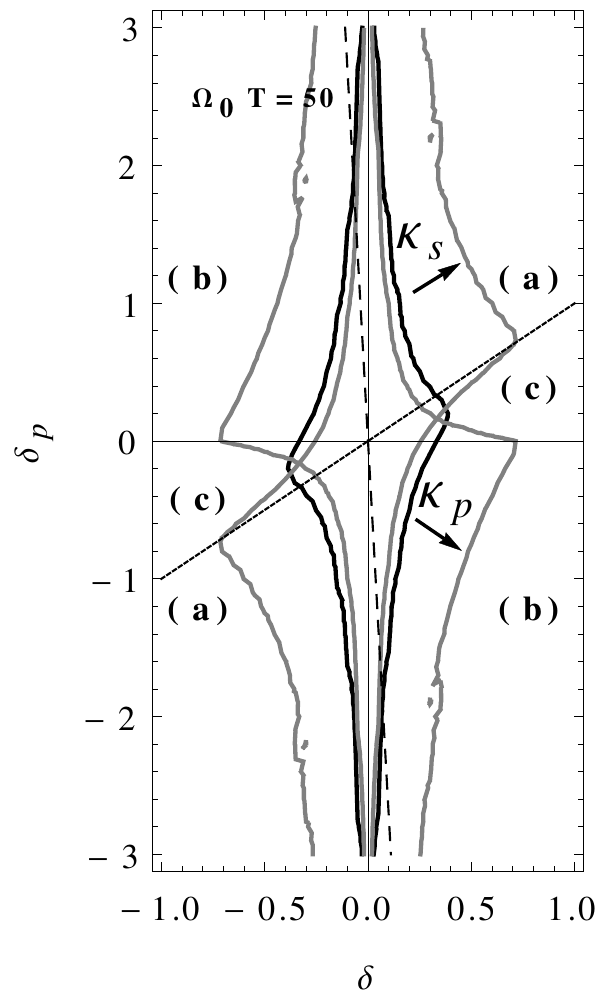}
\end{minipage}
%\end{minipage}
\caption{Left panel: typical Landau-Zener (LZ) patterns as a function 
of time of instantaneous eigenvalues for nonzero detunings. 
Three different patterns may occurr according to the value 
of $\delta_p/\delta$ (see Sec.\ref{sec:figure-merit}).  
Right panel: efficiency diagram of STIRAP vs 
detunings. The curves enclose regions corresponding to 
efficiency larger than 
90\%. The black line is obtained for 
$\kappa_p=\kappa_s$ and 
shows that the protocol is much more sensitive 
to deviations of $\delta$ rather than of $\delta_p$.
Gray lines are the efficiencies when $\kappa_p=2\kappa_s$ and $\kappa_s=2\kappa_p$
(see Sec.~\ref{sec:figure-merit}). 
Intercepts of such curves with lines $\delta_p/\delta = a$ 
define the two-photon linewidths $\delta_{\frac{1}{2}}$ as 
a function of all the parameters but $\delta$. The dashed line shows the example of the Cooper pair box for $q_g=0.48$ and $J=1$.
\label{fig:nonidealstirap}}
\end{figure}
Adiabaticity in ideal STIRAP~\cite{kr:198-bergmann-rmp-stirap,kr:201-vitanov-annurev} requires that $\Omega_0 T > 10$. 
Since it involves in a clever sequence several coherent 
phenomena~\cite{kr:201-vitanov-annurev} (AT effect, EIT and adiabatic 
passage),
STIRAP is very efficient, faithful and 
stable apart for a crucial sensitivity to $\delta$
(see Fig.~\ref{fig:nonidealstirap} right panel). Indeed  for 
$\delta \neq 0$ no exact dark state exists providing the
adiabatic connection $\ket{0} \to \ket{1}$. Still population 
transfer may take place 
by {\em non-ideal STIRAP},  
via non-adiabatic Landau-Zener (LZ) 
transitions between adiabatic states
(see Fig.~\ref{fig:nonidealstirap} left panel), a  
mechanism 
crucial for the applications in artificial atoms 
(Sec.\ref{sec:figure-merit}).

In artificial atoms the Hamiltonian 
reads~\cite{ka:214-paladino-rmp,ka:212-falci-physscr} 
$H = H_0({\mathbf q}) +
[{\cal A}_p(t) \cos(\omega_p t) + {\cal A}_s(t) \cos(\omega_s t) ] {\cal P}$, 
where the device $H_0$ depends on 
tunable parameters $\mathbf q$. The field
couples to the system operator $\cal P$ and  
the envelopes  ${\cal A}_k(t), k=p,s$ are slowly varying
with respect to Rabi frequencies. Under suitable conditions,
$H$ can be truncated to three levels.
Performing 
the RWA and transforming to a doubly 
rotating frame, we get the
form (\ref{eq:H}), 
where $\Omega_p(t)={\cal A}_p(t) \,{\cal P}_{20}$ and 
$\Omega_s(t)={\cal A}_s(t) \,{\cal P}_{21}$.

\section{Effective model for solid-state noise}
\label{sec:model-noise}
Physically, noise in solid-state devices has 
large low-frequency components with a $1/f^\alpha$ 
spectrum, and high frequency component, either white or 
ohmic. 
Assuming for simplicity a single noise source 
inducing fluctuations of the parameter $q_g$, 
we can describe this BBCN by the phenomenological 
Hamiltonian~\cite{ka:214-paladino-rmp}
$H = H_0(q_g+x(t)) + {\cal A}(t) {\cal P} + H_{env}$.
Here $x(t)$ is a classical stochastic process accounting for
low-frequency noise, whereas $H_{env}$ describes an environment 
coupled to the system, responsible for Markovian quantum noise.
\begin{comment}
We consider for simplicity a single noise source, 
inducing fluctuations of the parameter $q_g$, and seek for 
a phenomenological description of BBCN by first considering  
classical noise, modeled by 
adding a stochastic process, $q_g \to q_g + x(t)$.     
Then we split BBCN 
into parts slow and a fast with respect to the 
typical time scales of the system, 
$x(t) \to x(t) + x_f(t)$, and quantize the latter 
obtaining the phenomenological 
Hamiltonian~\cite{ka:212-falci-physscr}
\begin{equation}
H = H_0(q_g+x(t)) + {\cal A}(t) {\cal P} + H_{fn}
\end{equation}
Here $H_{fn}=\hat{X} {\cal Q} + 
H_{env}$ describes ``fast noise''  as due to an 
environment $H_{env}$, coupled to the system operator ${\cal Q}$
via an environment operator $\hat{X}$. 
\end{comment}
The effect of low-frequency noise is obtained by 
averaging over the stochastic process the density matrix 
$\rho^{f}(t| q_g+x(t))$, accounting for fast noise 
in a background stochastc field
\begin{equation}
\rho(t) = \int {\cal D} x(t) \, P[x(t)] \,\,
\rho^{f}(t| q_g+x(t)) 
\end{equation}
Leading effects are estimated by evaluating the integral in 
the ``quasistatic'' or static path approximation (SPA),  
i.e. by substituting 
$x(t)$ with a 
random variable $x$ with distribution $p(x)$ 
and calculating $\rho^{f}(t|q_g+x)$ by a Markovian 
master equation. 

Notice that $H_0$, its eigenenergies $E_i(q_g)$ 
and the matrix elements $P_{ij}$ entering 
$\Omega_{ij}(q_g)$ depend on $q_g$. 
A proper choice of $q_g$ may enforce 
symmetries of $H$, which protect the system against 
dephasing due to fluctuations of $E_i(q_g)$, but 
at the same time suppress some $P_{ij}$. 
Non-Markovian noise determines fluctuations of 
the entries of the Hamiltonian (\ref{eq:H}), namely   
$\delta_k(q_g+x) = \Delta E_k(q_g+x)-\omega_k$ and 
$\Omega_k(q_g+x) = 
{\cal A}_k(t)\,{\cal P}_k(q_g+x)$, 
where $\Delta E_k(q_g+x)$ and ${\cal P}_k(q_g+x)$ 
are the relevant energy splittings and "dipole" 
matrix elements.
This is a key issue for all our subsequent analysis
about design and optimization of Lambda systems. 
For instance, it is clear that for a Lambda configuration
at nominal resonance, i.e. if external fields 
are resonant at the nominal bias $q_g$, 
%$\delta(x) \to E_1(q_g+x) - E_0(q_g+x) - (\omega_p-\omega_s)$. Therefore 
fluctuations in the ``trapped subspace'' translate 
in stray $\delta \neq 0$ which are the most detrimental for 
STIRAP. It is convenient to expand detunings and Rabi frequencies. For instance at nominal resonance and 
for small enough fluctuations, imposing $E_0=0$ we have
\begin{equation}
\label{eq:delta-x}
\begin{aligned}
\delta(x) = 
A_1(\mathbf{q}) x + \frac{1}{2} B_1(\mathbf{q}) x^2
\quad ; \quad
\delta_p(x) = \Delta E_2 = 
A_2(\mathbf{q}) x + \frac{1}{2} B_2(\mathbf{q}) x^2
\end{aligned}
\end{equation}
where $A_i(\mathbf{q})=\partial E_i(\mathbf{q})/\partial q_g$ 
and $B_i(\mathbf{q})=\partial^2 E_i(\mathbf{q})/\partial q_g^2$. 
%The analogous expansion for fluctuations of the drives starts from the value ${\cal P}_{ij}(\mathbf{q})$ at $q_g$. 
We notice that
all such fluctuations are correlated via the bandstructure 
of the device, since they originate 
from the same random variable $x$.

%\subsection{A case study: the Cooper Pair Box}\label{sec:num-res}
We apply these ideas to the important case study of the 
Cooper pair box (CPB), a superconducting 
device described by the Hamiltonian  
\begin{equation}
\label{eq:CPB-H}
H_0(q_g,J) = \sum_n (q_g - n)^2 \ketbra{\phi_n}{\phi_n} - 
\frac{J}{2} (\ketbra{\phi_{n+1}}{\phi_n} + h.c.).
\end{equation}
Here $\hat{n}:=\sum_n n \ket{\phi_n}\bra{\phi_n}$ is the number 
of extra Cooper pairs 
in a metallic island. 
The parameter $J$ is the relative strength of the Josepson 
tunneling changing $n \to n \pm 1$.   
According to its value several different implementations 
of superconducting qubits from ``charge qubits'' to ``transmons''~\cite{ka:199-nakamura-nature,ka:204-duty-delsing-prb-chargeqb,ka:202-vion-science,ka:204-wallraff-nature-cqed,ka:207-koch-pra-transmon} have been demonstrated. 
The other parameter $q_g$ can be tuned by an external voltage. 
The CPB is operated by an ac gate voltage which is coupled 
to the charge, ${\cal P}=2 e \hat{n}$, playing 
the role of the system operator. 

The Hamiltonian $H_0(q_g,J) = \sum_i E_i \ket{i}\bra{i}$ 
is symmetric 
for charge parity transformations
at half-integer and integer $q_g$. 
Here the selection rule
$n_{ij} := \bra{i}\hat{n}\ket{j} = [1-(-1)^{i+j}] n_{ij}$ holds, preventing pump 
coupling.
On the other hand working at $q_g = 1/2$ guarantees
the maximal protection from charge 
noise because of the suppression of $A_i(\mathbf{q})$ in Eq.(\ref{eq:delta-x}). Larger values of $J \gg 1$ 
suppress asymmetries at $q_g \neq 1/2$, ensuring
protection in a larger region of the space 
of parameters, 
where however the pump coupling is suppressed~\cite{ka:213-falci-prb-stirapcpb}. 

\begin{figure}[t]
\centering
\includegraphics[width=0.42\columnwidth]{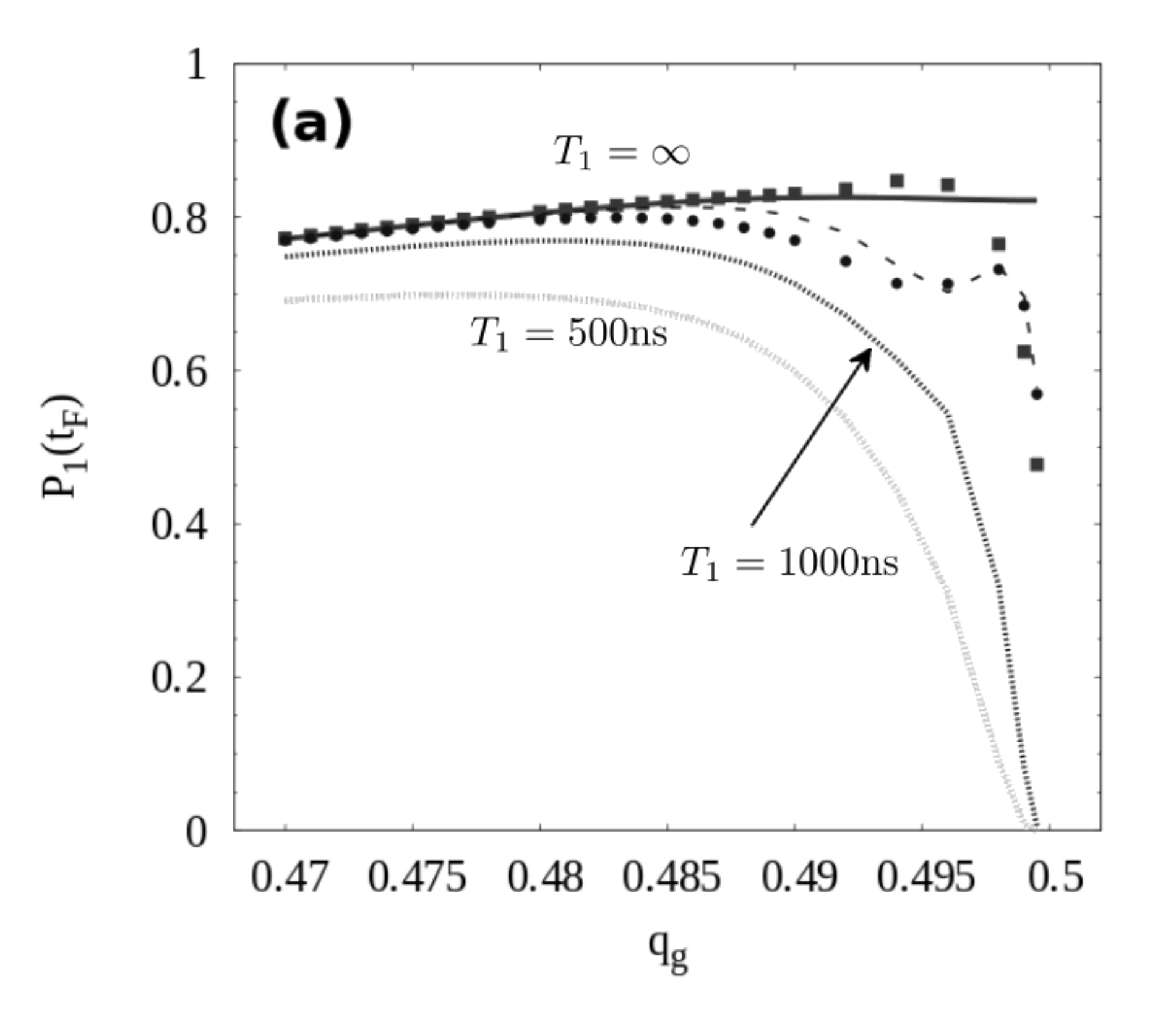}
\includegraphics[width=0.35\columnwidth]{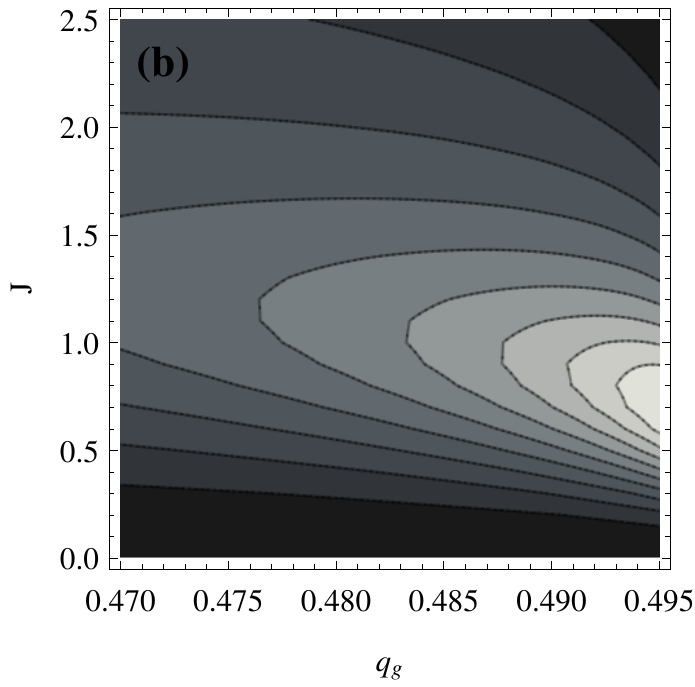}
\raisebox{10mm}{\includegraphics[width=0.07\columnwidth]{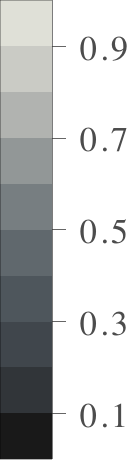}}
\caption{(a) Efficiency of STIRAP in the CPB 
for $J=1$, showing the 
effect of low-frequency noise alone and with additional Markovian noise.
Field amplitudes, corresponding to 
Rabi oscillations $\Omega_R = 2 \pi \cdot 600 \mathrm{rad/s}$ 
in the
Quantronium qubit~\cite{ka:205-ithier-prb} 
operated at the symmetry point, 
yield a sufficiently large pump frequency 
$\Omega_0 \simeq \Omega_R n_{02}(q_g)/n_{01}(0.5)$ for 
values of $q_g$ which slightly break the symmetry. 
Here we consider linear fluctuations of detunings (thick solid line),  
while adding their quadratic fluctuations (squares) and 
fluctuations of $\Omega_k$ in linear (dashed line) 
and quadratic approximation (dots) has no effect. 
Markovian noise determines a smaller further reduction of the efficiency (thin gray lines).
We used a variance $\sigma_x = 0.004$ of 
$p(x)$ and $T_1= 500,1000\,\mathrm{ns}$
(b) Figure of merit Eq.(\ref{eq:figure-merit}) 
for design and optimal symmetry breaking against 
low-frequency charge noise in the CPB. 
We consider the region where charge noise is linear since 
closer to the symmetry point $q_g = 1/2$, 
STIRAP is suppressed by sèpontaneous decay.
\label{fig:results-cpb}
}
\end{figure}

The numerically 
calculated~\cite{ka:213-falci-prb-stirapcpb} 
efficiency of STIRAP 
(Fig.~\ref{fig:results-cpb}a) shows that a Lambda configuration
allowing population transfer $\sim 80\%$ is 
achievable in a CPB with $J=1$ by operating at a symmetry 
breaking bias $q_g \neq 1/2$, despite of the reduced protection 
from low-frequency noise. In this regime, 
$q_g \lesssim 0.49$, only 
linear fluctuation of detunings matter, i.e. linear terms 
in Eqs.(\ref{eq:delta-x}) are considered.  
We used $\Omega_0 T=15$ to guarantee good adiabaticity, 
with figures of noise and couplings consistent with 
measurements of the decoherence in the qubit of 
Ref.~\cite{ka:205-ithier-prb}.
In this regime, Markovian emission~\cite{kr:198-bergmann-rmp-stirap,kr:201-vitanov-annurev} or absorption~\cite{ka:213-falci-prb-stirapcpb} channels are not effective, 
whereas spontaneous decay in the trapped subspace, 
characterized by $T_1$, is tolerably small. 
Instead near $q_g = 1/2$ pump coupling is small 
and it would require $T \gg T_1$, thereby decay suppresses
the efficiency. 

\section{Strategies of  protection against noise}
\label{sec:figure-merit}
We now analize the tradeoff between efficient coupling 
and decoherence when parity symmetry is broken.
Our analysis leverages on the results of last section, 
that the main mechanism of efficiency loss (besides 
lack of adiabaticity) are low-frequency fluctuations 
of energy levels. As discussed in 
Sec.\ref{sec:model-noise}, these 
correspond to fluctuations of detunings and can be 
represented on the diagram of Fig.~\ref{fig:nonidealstirap} 
by a curve. 
For vanishing nominal detunings
it has parametric equations (\ref{eq:delta-x})
and passes the origin for $x=0$. 
Notice that in general in artificial atoms 
fluctuations of detunings 
produced by a specific noise source 
show specific correlations.
In the linear noise regime 
$\delta_p/\delta = a$, where $a= A_2(\mathbf{q})/A_1(\mathbf{q})$, thereby the signs of the derivatives 
$A_i(\mathbf{q})$ of the bandstructure determine whether
detunings are correlated (e.g. critical current noise in flux qubits) or anticorrelated (e.g. charge noise in the 
CPB)~\cite{ka:214-paladino-rmp}.
It has been shown~\cite{ka:212-falci-physscr} 
that according to such 
correlations, three different kind of 
LZ patterns may occurr. They are shown in 
Fig.~\ref{fig:nonidealstirap} (left panel) and labelled with 
(a) for $\delta_p/\delta > 1$, (b) for 
$\delta_p/\delta < 0$ and (c) for $0\le \delta_p/\delta < 1$. 
The efficiency dependence on parameters turns out to be 
different in the three cases, this being the key 
to classify effects of low-frequency noise and the specific 
strategies to suppress them.

%\subsection{Optimal symmetry breaking strategy}
\label{sec:noise-detunings}
Since {\em both} couplings and fluctuations depend on 
the solution of the eigenproblem of $H_0(\mathbf{q})$,  
we may seek for an ``optimal'' 
set of values $\mathbf{q}$ such that the symmetry breaking 
yields enough pump coupling still keeping decoherence 
tolerable. Referring to Fig.~\ref{fig:nonidealstirap}   
%successful STIRAP requires that fluctuations $x \in [-\sigma_x,\sigma_x]$ should correspond to a line segment lying in the large efficiency region. We 
we formulate this condition by defining a  
{\em two-photon linewidth}~\cite{kr:201-vitanov-advmolopt} 
$\delta_{\frac{1}{2}}$, as 
the interval of $\delta$ where coherent transfer 
is appreciable, for any fixed combination of the other parameters
(see Fig.~\ref{fig:nonidealstirap}). For efficient STIRAP 
low-frequency noise must induce 
fluctuations of $\delta$ with small enough 
variance, 
$\sigma_{\delta} \approx \sqrt{A_1^2 \sigma_x^2  + 
{1 \over 2} B_1^2 \sigma_x^4} \lesssim \delta_{1/2}$.

The linewidth can be estimated by evaluating the 
impact of unwanted  
transitions between adiabatic states. In this way 
Vitanov et al.~\cite{kr:201-vitanov-advmolopt} 
found the scaling law 
$\delta_{\frac{1}{2}}\simeq d(\tau) \Omega_0
\sqrt{\kappa_p^2+\kappa_s^2}$
valid for $\delta_p = 0$, and 
roughly holding on the lines $\delta_p/\delta = a$ in 
the region (c) of Fig.~\ref{fig:nonidealstirap} left. 
In the same way one can derive that 
$\delta_{1 \over 2} \approx d^\prime(\tau,\kappa) \Omega_0
\,\kappa_p$ in the region (b) of anticorrelated 
detunings~\cite{ka:213-falci-prb-stirapcpb}, whereas 
$\delta_{1 \over 2} \approx d^{\prime\prime}(\tau,\kappa) \Omega_0 \kappa_s$ in the region (a) 
of correlated detunings.
The dependence on $\kappa =\kappa_s/\kappa_p$ in the 
prefactors turns out to be weak. 
\begin{comment}
\begin{figure}
\centering
\begin{minipage}[c]{0.7 \columnwidth}
\begin{minipage}[c]{0.85 \columnwidth}
\includegraphics[width=\columnwidth]{Images/meritFigure}
\end{minipage}
\begin{minipage}[c]{0.1 \columnwidth}
\includegraphics[width=\columnwidth]{Images/MeritFigureLegend}
\end{minipage}
\end{minipage}
\caption{Figure of merit Eq.(\ref{eq:figure-merit}) 
for design and optimal symmetry breaking against 
low-frequency charge noise in the CPB. 
We consider the region where charge noise is linear since 
closer to the symmetry point $q_g = 1/2$ 
STIRAP is suppressed by sèpontaneous decay.
\label{fig:figure-merit}
}
\end{figure}
\end{comment}

In the case of
CPB, since $A_1$ and $A_2$ have different sign, charge noise
determines anticorrelated fluctuations of detunings, and 
good transfer efficiencies are achieved 
for large values of the ratio
\begin{equation}
\label{eq:figure-merit}
{\delta_{1 \over 2} \over \sigma_\delta} \propto 
{\kappa_p\Omega_0 \over \sigma_{\delta}} \approx 
{2 \, n_{02}(J,q_g) 
\over \sqrt{A_1^2(J,q_g) \sigma_x^2 + {1 \over 2} B_1^2(J,q_g) \sigma_x^4}}.
\end{equation} 
This is a figure of merit for 
STIRAP efficiency %depending on the parameters $(J,q_g)$   
(see Fig.~\ref{fig:results-cpb}b) which can be used for  
seeking optimization of both the design of the device and 
the symmetry breaking of the Hamiltonian modulated on-chip 
by the bias $q_g$. 

%\subsection{Dynamical decoupling}
The above analysis also suggests 
that effects of charge noise in a CPB can be 
minimized by increasing $\kappa_p$ only. This 
is a specific way of decoupling dynamically 
noise sources, responsible for 
anticorrelated $(\delta,\delta_p)$ fluctuations.
Indeed it is clear from Fig.~\ref{fig:nonidealstirap}b 
(the blue curve) that increasing $\kappa_p$ the 
efficiency grows in the region (b), suppressing 
anticorrelated fluctuations of detunings. 
This happens because non ideal STIRAP occurs via LZ 
tunneling along the pattern 
(b) in Fig.~\ref{fig:nonidealstirap}, being suppressed 
for increasing $\delta$ when the avoided crossing builts on, 
and being restored if the gap shrinks due to a  
larger $\Omega_p$. 

This analysis can be extended to the 
main different designs of superconducting artificial atoms,
and to each specific low-frequency noise source. 
These latter are classified according to 
the $(\delta,\delta_p)$ correlations they determine. 
For instance flux noise in flux qubits yields 
anticorrelated $(\delta,\delta_p)$, 
as for charge noise in the CPB, and 
increasing $\Omega_p$ yields dynamical decoupling.
Instead critical current noise in CPB and flux qubit 
determine correlated
$(\delta,\delta_p)$ fluctuations, requiring larger $\Omega_s$.
In phase qubits both critical current and bias current noise 
yield correlated $(\delta,\delta_p)$ fluctuations dynamically suppressed by a large $\Omega_s$. 

%\subsection{Applications to artificial atoms}
In real superconducting artificial atoms, where 
more than one noise source is present, the two strategies
can be combined. Protection from noise producing 
anticorrelated $(\delta,\delta_p)$ fluctuations can be achieved  
by the optimal symmetry breaking strategy, since  
dynamical decoupling is limited by the 
insufficient coupling $\Omega_p$. Protection 
from  noise producing correlated $(\delta,\delta_p)$ 
fluctuations can then be obtained increasing $\Omega_s$,
which is not limited by selection rules.

It is easy to extend this analysis to artificial atoms driven in 
different field configurations. For instance for population 
transfer in the Ladder scheme(Fig.~\ref{fig:stirap-ideal})
scheme one associates $\delta$ ($\delta_p$) with the second 
(first) excited state, which allows to identify
the relevant correlations between detunings.

\section{Implications of non-Markovianity}
\label{sec:non-markovianity}
The picture of the last section relies on the non-Markovianity 
of BBCN. We remind that 
low-frequency noise is the main source 
of dephasing in artificial atoms. BBCN explains distinctive 
experimental observations in quantum bits~\cite{ka:214-paladino-rmp,ka:205-ithier-prb,ka:211-bylander-natphys,ka:212-chiarello-njp}. Moreover design of low-decoherence qubits relies on 
protection from non-Markovian noise. Both optimal tuning~\cite{ka:211-paladino-njp,ka:210-paladino-prb}
and dynamical decoupling~\cite{ka:212-lofranco-physscr-entdyn,ka:214-lofranco-prb-entdecoup} 
have been exploited for 
entangled states
We generalize this ideas 
to protection of three-level coherence, obtaining a 
rich scenario. 

It is important to point out the different impact on STIRAP 
of non-Markovian dephasing, as discussed in this work, 
and Markovian pure dephasing as described by the 
standard Master Equation approach. This latter 
problem has been studied in 
Ref.~\cite{ka:204-ivanov-pra-stirapdephasing},
including only the dephasing rates 
$\tilde{\gamma}_{ij}$. For large enough $\Omega_0T$ 
populations at the end of 
the protocol were found to be
\begin{equation}
\label{eq:mark-rho-infty}
\begin{aligned}
\rho_{11}(\infty) =\frac{1}{3}+\frac{2}{3}\mathrm{e}^{-3\tilde{\gamma_{01}}T^2/8\tau}\quad ; \quad
\rho_{00}(\infty) = \rho_{22}(\infty) =\frac{1}{3}-\frac{1}{3}\mathrm{e}^{-3\tilde{\gamma_{01}}T^2/8\tau}
\end{aligned}
\end{equation}
i.e. dephasing determines efficiency losses which 
do not depend on the peak Rabi frequencies. Therefore 
Markovian dephasing cannot at all accout for the scenario 
presented in Sec.~\ref{sec:figure-merit}.

\begin{figure}
\centering
\includegraphics[width=0.4\columnwidth]{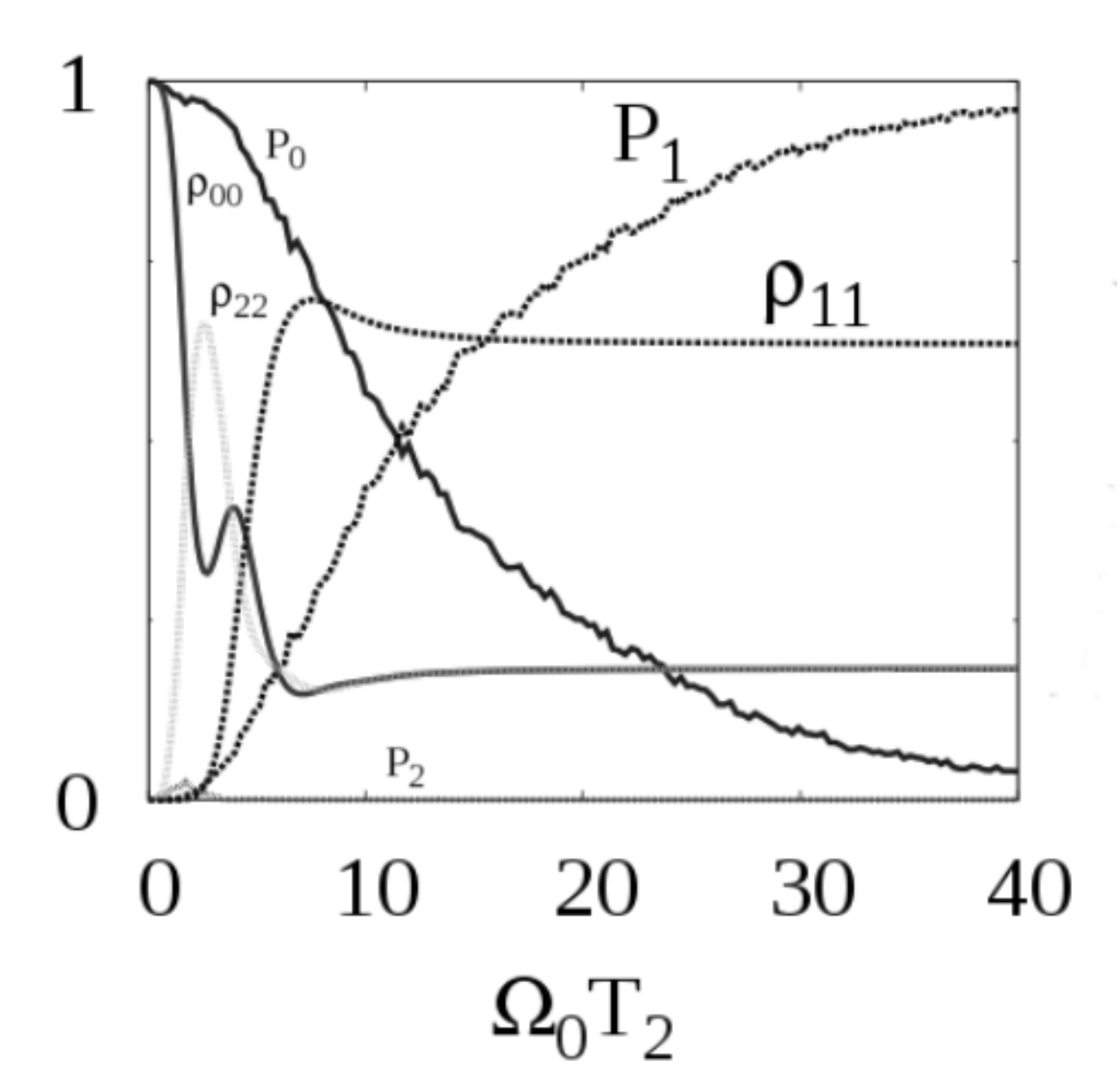}
\caption{Final populations of STIRAP 
with Markovian ($\rho_{ii}$) and non Markovian ($P_i$) noise. 
The former is the solution of a Master Equation with 
rate $\tilde{\gamma}_{10}= 1/T_2$, leading to 
exponential decay of qubit coherences. The non-Markovian 
noise is simulated taking a distribution of detunings 
corresponding to $\sigma_x= \sqrt{2}/(A_1 T_2)$,
leading to Gaussian decay with the same $T_2$. 
\label{fig:dephasing}
}
\end{figure}

In Fig. \ref{fig:dephasing} we plot the final populations of 
the bare states comparing Markovian ($\rho_{ii}$) 
and non-Markovian ($P_i$) pure dephasing, in the entire 
relevant range of $\Omega_0$. Noise produces in both cases 
the same qubit dephasing time $T_2$,
which is relatively large. 
For large $\Omega_0$ while for non-Markovian noise 
$P_1$ is completely recovered, for Markovian noise it  
saturates to a smaller value  given by 
Eq.(\ref{eq:mark-rho-infty}). 
For small $\Omega_0$ the protocol fails in both cases, 
due to insufficient adiabaticity. 

\section{Conclusions}
\label{sec:conclusions}
In this work we discussed effects of BBCN noise in three level 
artificial atoms. In particular we studied the tradeoff between 
protection from low-frequency noise, enforced by symmetries of 
the Hamiltonian, and the implied selection rules which are  
the main obstacle to the implementation of a Lambda scheme. 
Being based on two-photon absorption {\em and} emission, the 
Lambda scheme allows performing tasks as 
transduction of photons in 
the $\mathrm{\mu m}$/mm range. We have studied STIRAP since 
it is a benchmark advanced protocol. It is also the basis 
of other protocols from preparation of 
superpositions~\cite{kr:201-vitanov-annurev}
to transfer of wavepackets
and manipulation of photons, with still unexplored 
potentialities for quantum information and quantum control. 

We have shown that model for BBCN noise decoherence in 
the ``trapped subspace'' $\mathrm{span}\{\ket{0},\ket{1}\}$
plays a major role, a conclusion which holds for all Lambda, 
Ladder and Vee schemes. Strategies to defeat noise in qubits 
can then be generalized to three-level systems. We presented 
two stetegies, namely optimal symmetry breaking and 
continuous dynamical decoupling, which can be integrated to 
minimize the effects of anticorrelated and correlated 
parametric fluctuations of the artificial atom bandstructure. 
Relying on non-Markovianity of BBCN, our results suggest that 
features of the scenario of STIRAP with BBCN, as the 
predictions on the peculiar dependence on control knobs
described in Sec.~\ref{sec:figure-merit}, could be used to probe aspects of non-Markovianity of the solid-state evironment.

Finally, we mention that artificial atoms allow for new 
unconventional schemes to manipulate a Lambda system, 
bypassing hardware constraints and allowing to perform STIRAP 
at protected symmetry points and with always-on fields.  
The strategies to defeat noise presented here could be 
successfully applied also in these cases.

\begin{acknowledgement}
This work was supported by MIUR through Grant. 
``ENERGETIC -- Tecnologie per l'ENERGia e 
l'Efficienza energETICa'', 
No. PON02\_00355\_3391233.
A.D.A. acknowledges 
support from Centro Siciliano di Fisica 
Nucleare e Struttura della Materia under grant MEDNETNA.
\end{acknowledgement}

\bibliography{stirap,misc}{}
\end{document}